# Research on the Online Update Method for Retrieval-Augmented Generation (RAG) Model with Incremental Learning

Yuxin Fan*
School of Engineering and Applied Science
University of Pennsylvania
Toronto, Canada
* Corresponding author e-mail: yuxinfan@alumni.upenn.edu

Lipeng Liu
College of Engineering
Peking University
Beijing, China
E-mail address: lipeng.liu@pku.edu.cn

Na Sun
Independent Researcher
Newark, USA
E-mail address: sunna825825@gmail.com

Yuxiang Wang
School of Business
Stevens Institute of Technology
New York, USA
E-mail address: yuxiang.wang0476@gmail.com

Xirui Tang
College of Computer Sciences
Northeastern University
Boston, USA
E-mail address: tang.xir@northeastern.edu

Zidong Yu
Syracuse University
College of Engineering and Computer Science
San Jose, USA
E-mail address: zidong.yu101@gmail.com

*Abstract*—In the contemporary context of rapid advancements in information technology and the exponential growth of data volume, language models are confronted with significant challenges in effectively navigating the dynamic and ever-evolving information landscape to update and adapt to novel knowledge in real time. In this work, an online update method is proposed, which is based on the existing Retrieval Enhanced Generation (RAG) model with multiple innovation mechanisms. Firstly, the dynamic memory is used to capture the emerging data samples, and then gradually integrate them into the core model through a tunable knowledge distillation strategy. At the same time, hierarchical indexing and multi-layer gating mechanism are introduced into the retrieval module to ensure that the retrieved content is more targeted and accurate. Finally, a multi-stage network structure is established for different types of inputs in the generation stage, and cross-attention matching and screening are carried out on the intermediate representations of each stage to ensure the effective integration and iterative update of new and old knowledge. Experimental results show that the proposed method is better than the existing mainstream comparison models in terms of knowledge retention and inference accuracy.

I. INTRODUCTION

The rapid development of information technology and the widespread use of the Internet have resulted in a considerable increase in the generation and accumulation of data on a global scale. This phenomenon has not only led to a substantial augmentation in the available research resources within the domain of natural language processing (NLP), but also ushered in unparalleled opportunities for its practical applications. However, the advent of massive data sets has introduced challenges of a more complex and varied nature to NLP. Language models, as the core technology of NLP, have demonstrated remarkable efficacy and extensive application potential in various domains, including machine translation, intelligent question answering, and text generation. To illustrate, machine translation systems employ deep learning models to achieve high-quality conversion between multiple languages, intelligent question answering systems can understand and answer complex questions from users, and text generation technologies play an important role in content creation and automatic summarization [1]. However, in the face of dynamic changes in the information environment, traditional static language models often prove difficult to adapt to new knowledge in real time, resulting in lagging model performance and outdated knowledge [2].

The Retrieval-Augmented Generation (RAG) model is an emerging language generation framework that has the capacity to access the information of an external knowledge base in real time during the generation process. This is achieved by combining a retrieval module and a generation module, thereby significantly improving the accuracy and richness of the generated content. RAG models have been shown to excel in the domains of open-domain question-and-answer and knowledge-intensive tasks. This is achieved by means of dynamic document retrieval and integration into the generation process, resulting in generated texts that exhibit higher factual accuracy and a broader range of knowledge domains [3]. For instance, in the context of open-domain Q&A, the RAG model has the capacity to retrieve relevant literature in real time and generate responses based on the latest information, thereby significantly enhancing the practicability and reliability of the system. Nevertheless, despite the RAG model's proficiency in handling static knowledge bases, numerous challenges persist in the realm of online updating and knowledge integration within a dynamic environment. The management of dynamic knowledge bases, the efficiency of real-time retrieval, and the effective integration of new and old knowledge are all key problems that need to be solved urgently, and the existence of these problems limits the promotion and application of RAG models in a wider range of practical application scenarios [4].

The highly dynamic nature of the information environment gives rise to the constant emergence of new knowledge and data, which calls for language models that are able to capture and integrate this new information in a timely manner, thereby maintaining their competitiveness in knowledge updates and application scenarios [5]. The ability of language models to swiftly adapt to novel knowledge in a perpetually evolving environment has emerged as a pivotal concern, with the objective being to enhance the practicality and intelligence of the model. Conventional model update methodologies frequently necessitate the retraining of the entire model, a process that is not only time-consuming and labour-intensive, but also often impractical in real-time scenarios [6]. Furthermore, the retraining process is susceptible to "catastrophic forgetting", a phenomenon in which the model disregards previously acquired knowledge while acquiring new information. This phenomenon severely restricts the performance of the model in real-time applications, making it difficult for the model to maintain stable and efficient performance in a rapidly changing environment.

## II. EASE OF USE

Fu et al. [7] proposed an automated online hyperparameter tuning method specifically designed to improve the performance of Retrieval Enhancement Generation (RAG) models. Conventional RAG models frequently employ static model design and fixed hyperparameter configuration. However, in view of the dynamic change of the information environment, static design often proves challenging in terms of coping with new situations in practical applications. The innovation of AutoRAG-HP lies in its ability to adjust hyperparameters in real-time, enabling the RAG model to adapt to the unique characteristics of the current data during the generation process. He et al. [8] proposed the Chain-of-Verification method to enhance the performance of RAG models. The purpose of this method is to enhance the accuracy and robustness of the generated content by incorporating multiple rounds of retrieval and generating feedback. Specifically, this paper proposes a unified enhanced generative model framework, which introduces multiple steps in the model generation process, such as preliminary retrieval, content generation, and final validation. The model's retrieval results are reversed at each step, allowing for continuous correction and improvement of the generated results.

Yue et al. [9] proposed an evidence-driven RAG response generation method specifically designed to address the problem of misleading information online. In this study, the authors propose a training strategy that combines reinforcement learning and human feedback (RLHF) to enhance the model's capacity to verify facts during the generation process, particularly in the generation and filtration of online disinformation. The RAG model proposed by Lewis et al. [10] is one of the pioneering studies of the Retrieval Enhanced Generation (RAG) framework, which has greatly promoted the development of retrieval-based generative tasks.

Furthermore, Mozharovskii [11] evaluated the ability of Retrieval Enhanced Generation (RAG) technology to handle programming tasks in an augmented learning management system (LMS). With the increasing popularity of programming education and the emergence of online learning platforms, the necessity to provide students with immediate and accurate feedback in a large-scale online learning environment has become a significant research direction. Radeva et al. [12] proposed a web application based on RAG technology, implementing and testing it. The application's objective is to furnish users with knowledge retrieval-based generation services by integrating RAG technology, particularly in complex technical problem-solving and knowledge graph generation.

## III. METHODOLOGIES

### A. Dynamic Memory and Retrieval Module

Dynamic memory banks (DMBs) are utilised for the storage and management of emerging data samples. The memory is configured with $M_t$ memory cells at time step $t$, with the representation vector of each memory cell $m_i$ defined as $m_i \in \mathbb{R}^d$. Upon the arrival of a new data sample $x_t$, its representation vector is initially generated by the encoder $E$, as illustrated in Equation 1:

$$m_t = E(x_t). \#(1)$$

In order to ensure the maintenance of a dynamic and finite memory, a sliding window mechanism is employed to impose a maximum capacity of $N$ on the memory. The subsequent Equations 2 detail the rules that govern the updating of the memory:

$$\mathcal{M}_t = \begin{cases} \mathcal{M}_{t-1} \cup \{m_t\} & if |\mathcal{M}_{t-1}| < N \\ (\mathcal{M}_{t-1} \setminus \{m_{old}\}) \cup \{m_t\} & otherwise \end{cases}, \#(2)$$

where the memory unit known as $m_{old}$ is considered to represent the earliest memory unit added to the memory bank. This mechanism ensures that the memory can dynamically adapt to new data while avoiding memory explosions.

In order to effectively integrate new knowledge into Core Model $F$, a Knowledge Displacement (KD) strategy is employed. The teacher model, designated $F'$, incorporates the most recent knowledge information. The objective of knowledge distillation is to minimise the discrepancy between the outputs of the core model and the teacher model. This is accomplished through the utilisation of a loss function, as outlined in Equation 3:

$$\mathcal{L}_{KD} = \sum_{i=1}^{B} KL\left(\sigma\left(\frac{F(x_i)}{\tau}\right) \parallel \sigma\left(\frac{F'(x_i)}{\tau}\right)\right), \#(3)$$

the batch size is denoted by $B$, the softmax function by $\sigma$, and the temperature parameter by $\tau$, which is employed to regulate the smoothness of the soft target. The incorporation of the temperature parameter $\tau$ facilitates the modulation of the degree of influence exerted by the teacher model on the core model, thereby circumventing the occurrence of overfitting with respect to the novel knowledge. When combined with Cross-Entropy Loss ($\mathcal{L}_{CE}$), the overall loss function is defined as shown in Equation 4:

$$\mathcal{L} = \alpha \mathcal{L}_{CE} + \beta \mathcal{L}_{KD}, \#(4)$$

In this model, $\alpha$ and $\beta$ represent weight parameters that are utilised to balance the classification loss and the distillation loss. By adjusting $\beta$, the impact of new knowledge on the core model can be flexibly controlled, ensuring the stability and knowledge integration effect in the online update process.

*B. Generation Modules*

The generation module is responsible for generating the final text output based on the retrieved contextual information. In order to ensure the effective integration and iterative update of new and old knowledge, a multi-stage network structure (MNS) and a cross-attention mechanism (CAM) have been designed, and the synchronous update and knowledge fusion of the model have been realised through a joint optimization strategy.

In the generation phase, a multi-stage network structure with $S$ generation stages was designed. Each stage $s$ contains a generation subnet $G_s$. The input vector $z$ goes through the first stage to generate a preliminary output $y_1$, which is then passed to the next stage for further processing, and so on, resulting in the final output $y_s$, expressed in Equation 5:

$$y_s = G_s(y_{s-1}, z), \#(5)$$

where $y_0 = z$. At each generation stage, the Intersecting Attention Mechanism (CAM) is introduced to match and screen intermediate representations of different stages. The calculation of cross-attention is outlined in Equation 6:

$$Attention(Q, K, V) = softmax\left(\frac{QK^T}{\sqrt{d_k}}\right)V, \#(6)$$

It is evident that among them, the query $Q$, key $K$, and value $V$ are intermediate representations from different stages. Through this mechanism, it is possible to interact and fuse information between different generation stages with each other, thereby ensuring the consistency and accuracy of the generated results.

In order to realise the synchronous update of the retrieval and generation modules, the overall loss function is defined and the optimisation processes of the two modules combined, as illustrated in Equation 7:

$$\mathcal{L}_{total} = \mathcal{L}_{retrieval} + \lambda \mathcal{L}_{generation}, \#(7)$$

$\mathcal{L}_{retrieval}$ is defined as the loss function of the retrieval module, and the negative log-likelihood of the retrieval accuracy is generally employed. The loss function of the generation module, termed $\mathcal{L}_{generation}$, combines cross-entropy loss and knowledge distillation loss. The weight parameter, denoted by $\lambda$, serves to balance the loss of retrieval and generation. The loss function of the generation module is defined as Equation 8:

$$\mathcal{L}_{generation} = \mathcal{L}_{CE} + \beta \mathcal{L}_{KD}. \#(8)$$

By employing joint optimisation, it is possible to ensure that the retrieval module provides high-quality contextual information, which in turn can be used effectively by the generation module for the purpose of text generation. This process of seamless integration and selection, as well as the updating of new and old knowledge, is thus achieved. The optimization process employs the backpropagation algorithm to simultaneously update the parameters of the retrieval module and the cattle module, as illustrated in Equation 9:

$$\theta \leftarrow \theta - \eta \nabla_\theta \mathcal{L}_{total}, \#(9)$$

where the letter $\theta$ represents all the parameters, and the letter $\eta$ represents the learning rate.

IV. EXPERIMENTS

*A. Experimental Setup*

In order to comprehensively evaluate the effectiveness of the proposed online update method in a dynamic environment, the Natural Questions (NQ) dataset from the real world was selected as the main test set. Constructed by the Google Research team, Natural Questions is predicated on the queries of actual users of the Google search engine, encompassing a plethora of domains of knowledge including science, history, geography, technology, and more. The dataset is distinguished by its substantial size, heterogeneity, and intricate complexity. The experimental procedure involved the preliminary processing, cleansing, and formatting of the questions and answers from the NQ dataset. Additionally, an external knowledge base was constructed utilising the long answer section to bolster the retrieval module of the RAG model. The dynamic knowledge update environment was simulated, and the sliding window mechanism was employed to manage the memory, thereby ensuring that the model could integrate new knowledge in a timely manner while retaining existing knowledge.

*A. Experimental Analysis*

In order to comprehensively evaluate the superiority of the proposed online update method, a variety of existing Retrieval Enhancement Generation (RAG) models were selected as the comparison baseline, including the traditional RAG model, AutoRAG-HP, Chain-of-Verification method, and Evidence-Driven Retrieval Augmented Response Generation (ERAG) method. As demonstrated in Figure 1, the proposed online update method is shown to be superior to the traditional RAG model, AutoRAG-HP, Chain-of-Verification and ERAG comparison methods in all evaluation indexes. In terms of generating accuracy, the scores of the proposed method increased by approximately 5%, 4% and 3%, respectively, indicating that the proposed method has significant advantages in the accuracy and relevance of generated content.

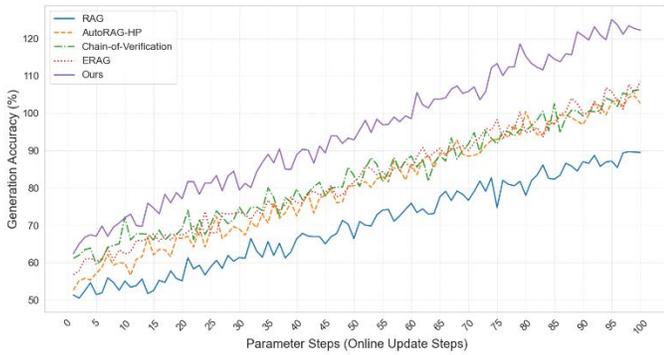

Figure 1.    Generation Accuracy Comparison Across Methods

Figure 2 shows a clear comparison of the differences in the performance of different approaches in terms of knowledge retention. Specifically, the proposed method "Ours" is significantly better than the traditional RAG model, AutoRAG-HP, Chain-of-Verification and ERAG comparison methods in the two indicators of non-forgetting rate and confusion test accuracy. This shows that our method can better retain and apply existing knowledge while effectively integrating new knowledge, avoid catastrophic forgetting, and improve the adaptability and stability of the model in a dynamic environment.

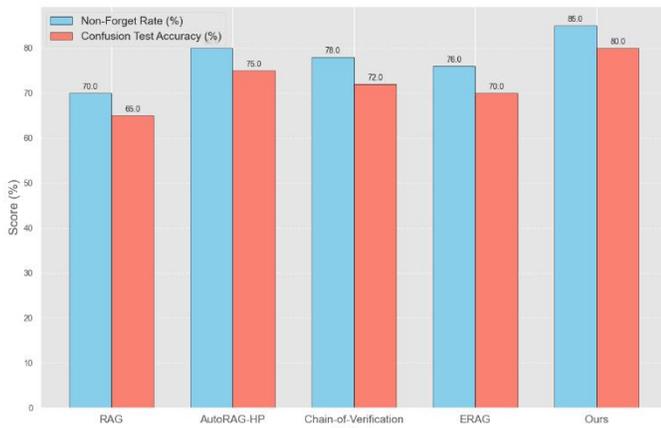

Figure 2.    Knowledge Retention Comparison Across Methods

In order to provide a visual illustration of the performance of different methods on Generative Consistency, Figure 3 compares the Generative Consistency scores of the traditional RAG model, AutoRAG-HP, Chain-of-Verification, ERAG, and the proposed method "Ours". The experimental results demonstrate that the proposed online update method, designated as "Ours", exhibits a significantly superior performance in terms of mass consistency when compared with the traditional RAG model, AutoRAG-HP, Chain-of-Verification and ERAG comparison methods. Specifically, the "Ours" method achieved a consistency score of 88.0%, which was significantly higher than that of other methods, with RAG of 75.0%, AutoRAG-HP of 82.5%, Chain-of-Verification of 80.0%, and ERAG of 78.0%. The error bars demonstrate that the method fluctuates less in the consensus score, indicating that it is more robust and reliable.

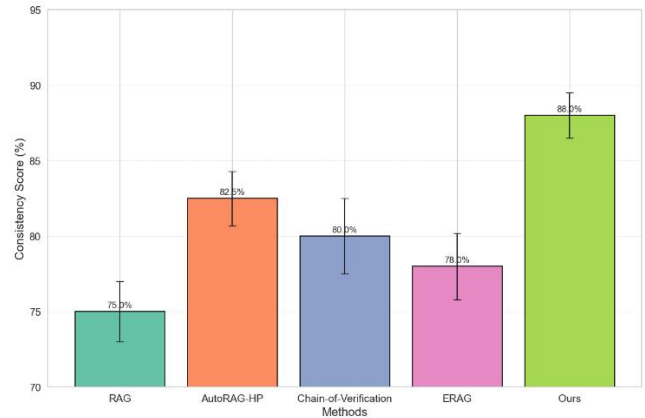

Figure 3.    Consistency Score Comparison Across Methods

## V. Conclusions

In conclusion, an innovative online update method combined with an incremental learning mechanism is proposed to address the challenges of online update and knowledge integration of retrieval-augmented generation (RAG) model in a dynamic environment. The model incorporates dynamic memory, a tunable knowledge distillation strategy, hierarchical indexing and multi-layer gating mechanism, multi-stage generative network structure and cross-attention mechanism. These innovations enable the model to retain and consolidate existing knowledge while adapting to new knowledge in real time. This enhances the performance and stability of the RAG model in dynamic knowledge environments. In the future, it is anticipated that the application effect of this method in various dynamic environments will be further enhanced through continuous optimisation and expansion, thereby making significant contributions to the intelligent and real-time development of natural language processing.